\documentclass[acmlarge]{acmart}
\makeatletter
\newcommand{\confshort}{\acmConference@shortname}
\newcommand{\conffull}{\acmConference@name}
\newcommand{\confdate}{\acmConference@date}
\newcommand{\confloc}{\acmConference@venue}
\AtBeginDocument{
  \fancypagestyle{firstpagestyle}{
    \fancyhead{}%
    \fancyfoot[C]{}%
  }
  \fancyhf{}
  \fancyhead[LO]{\@headfootfont\shorttitle}%
  \fancyhead[RE]{\@headfootfont\@shortauthors}%
  \fancyhead[LE]{\@headfootfont\footnotesize \confshort, \confdate, \confloc}%
  \fancyhead[RO]{\@headfootfont\footnotesize \confshort, \confdate, \confloc}%
  \fancyfoot[C]{}%
}
\makeatother
\acmBooktitle{\conffull\@ (\confshort), \confdate, \confloc}

\usepackage{tabularx} 

\AtBeginDocument{%
  }

\setcopyright{acmlicensed}
\copyrightyear{2026}
\acmYear{2026}
\setcopyright{cc}
\setcctype{by}
\acmConference[FAccT '26]{The 2026 ACM Conference on Fairness, Accountability, and Transparency}{June 25--28, 2026}{Montreal, QC, Canada}
\acmBooktitle{The 2026 ACM Conference on Fairness, Accountability, and Transparency (FAccT '26), June 25--28, 2026, Montreal, QC, Canada}
\acmDOI{10.1145/3805689.3812389}
\acmISBN{979-8-4007-2596-8/2026/06}

\begin{document}

\title[Overreliance in Writing Tasks]{Overreliance in Writing Tasks: Exploring Similarity-Based Measures of AI Influence on Writing and Proposing a Reflective Writing Interface Intervention}

\author{Vitor H. A. Welzel}
\affiliation{%
  \institution{Simon Fraser University}
  \city{Burnaby}
  \country{Canada}}
\email{vhd1@sfu.ca}
\renewcommand{\shortauthors}{V Welzel}

\author{Nicholas Vincent}
\affiliation{%
  \institution{Simon Fraser University}
  \city{Burnaby}
  \country{Canada}}
\email{nvincent@sfu.ca}
\renewcommand{\shortauthors}{N Vincent}

\renewcommand{\shortauthors}{Welzel and Vincent}

\begin{abstract}
As generative AI (GenAI) systems become increasingly proficient at simulating human-like and well-reasoned text, users may attribute authority to AI outputs, shaping how they engage with writing and reasoning tasks. While prior work has raised concerns about \textit{AI overreliance}, empirical approaches for observing this phenomenon during open-ended writing remain limited. In this paper, we examine how GenAI assistance influences users’ interactions with AI suggestions during writing. We report results from a mixed-methods study in which 47 participants completed analysis and synthesis writing tasks with or without AI assistance. We quantify the textual overlap between AI suggestions and participants' writing and analyze participants' reflections. Our results show that AI assistance is associated with patterns of suggestion reuse. Building on these findings, we design and evaluate an interactive writing interface that may support reflection on the usage of the AI suggestions during writing. Evidence from a small follow-up think-aloud study ($n = 4$) suggests that the interface can increase users’ awareness of how AI outputs are incorporated into their writing and may support more conscious engagement with AI assistance. Together, our findings contribute empirical methods for studying AI adoption in writing contexts and demonstrate how interface design can shape user–AI interaction.
\end{abstract}



\maketitle

\section{Introduction}
Generative AI (GenAI) systems are increasingly capable of producing text that reflects human reasoning cues, including a consistent natural tone and textual structure. Users may interpret these cues as indicators of credibility and expertise. Particularly in low‑trust environments, people rely on these mental proofs to assess credibility and sincerity \cite{wojtowicz2025underminingmentalproofai}. Previous work has shown that AI systems can simulate these proofs by crafting convincing arguments that users mistake for deep cognitive depth, therefore amplifying perceived expertise and trust \cite{danry2024deceptiveaisystemsexplanations}. Understanding how these perceptions shape user interaction with AI is an important challenge for human-computer interaction research.

A growing research area highlights the tendency of \textit{overreliance} on AI suggestions, accepting them even when they are incorrect, which can lead to poorer decision outcomes than independent reasoning \cite{vasconcelos2023explanationsreduceoverrelianceai, Bu_inca_2021, Kim_2025, bansal2021doesexceedpartseffect}. Interface interventions, such as explainable AI \cite{vasconcelos2023explanationsreduceoverrelianceai, Zhang_2020}, cognitive forcing functions \cite{Bu_inca_2021}, and clickable source references \cite{Kim_2025}, have shown some success in reducing overreliance, but often at the cost of usability or user satisfaction \cite{Bu_inca_2021}. 

Much of the previous research has focused on binary decision-making tasks and has overlooked the impact of AI on open-ended tasks. Open-ended tasks generally require analysis and synthesis skills that are vital for human cognitive development. Unlike binary decision-making, which often yields clear answers, open-ended tasks require deeper thinking and understanding, resulting in more nuanced responses. They involve interpreting various factors, such as creating solutions that may not be straightforward. This nuance is crucial for developing higher-order cognitive skills, according to Bloom's taxonomy \cite{bloom1956taxonomy, Hui07022025, singh2025protectinghumancognitionage}. Frequent usage of AI has also been linked to cognitive offloading, shifting users from "doing" to "overseeing" the AI work \cite{10.1145/3706598.3713778} as well as weakening neural indicators of mental engagement \cite{kosmyna2025brainchatgptaccumulationcognitive}. This shift may reduce opportunities to develop higher-order skills \cite{10.1145/3706598.3713778, kosmyna2025brainchatgptaccumulationcognitive, AIwriting, driverassistance}.

Through an experiment, we investigate how users incorporate AI suggestions into their responses to open-ended tasks and further analyze their behaviour after completing each task. Building on the idea of \textit{overreliance}, we provide additional evidence that people likely experience some degree of  AI influence, i.e. the adoption of wording, structure, and sentiment. This is not necessarily a concern in some contexts — perhaps producing outputs that are similar to high-quality AI outputs is beneficial. Specifically, in a small experiment, we measure the effects of AI suggestions on textual similarity, sentiment alignment, and perceived cognitive load. Our findings illustrate the transition from more noticeable to more subtle yet meaningful ways in which generative AI can influence users' responses.

In general, we might expect that as people see AI suggestions, their outputs become more similar to the AI outputs. In some cases, this may be the user's intention. In other cases, it might be cause for concern for the user or other people affected by the content. In a writing context, even seeing an AI suggestion might warp a creator's eventual output. 

Therefore, the open question, and our focus, is what happens when users do \emph{not} copy verbatim: to what extent do suggestions still imprint wording, structure, and sentiment on open-ended writing. To answer this, we argue that a single metric is insufficient; influence is not monolithic, and can manifest itself as lexical, structural, or tonal alignment. 

In this paper, we use the term \textit{adoption} to refer to \textbf{the incorporation of AI output into a participant's written response}, rather than to the general adoption of AI tools (e.g., increased ChatGPT usage). In practical terms, adoption refers to the extent to which a participant's final text aligns with the provided suggestion in wording, structure, meaning, and sentiment. We use \textit{overreliance} to refer to the broader concern from prior work. We discuss the main differences and when adoption becomes a concern in the Discussion section.

We find that \textbf{participants exposed to the fully formed AI suggestion exhibit higher levels of AI adoption}. This adoption is reflected in the increase in textual and sentiment similarity metrics. Building on these findings, we designed and implemented an interactive writing interface that displays real-time metrics to users on AI response adoption. We evaluate this interface through a follow-up, exploratory think-aloud study, which shows that reflection on such metrics can increase users' awareness of suggestion reuse.

Our work produces evidence and design practices for LLM tools:
\begin{itemize}
  \item We conducted an empirical study with participants to assess whether viewing AI output leads users to produce more similar output.
  \item We identified similarity metrics that may be particularly affected by AI assistance.
  \item We design and evaluate an interface intervention that can promote reflective engagement with AI responses in writing workflows.
\end{itemize}

\section{Related Work}
\subsection{Mental Proofs}
\citet{wojtowicz2025underminingmentalproofai} describe \textit{mental proofs} as a set of behaviours and cues that enable people to externalize their internal cognitive states \cite{understandingandsharingintentions}, such as understanding, sincerity and intention \cite{harland2024aiapologycriticalreview}. In low-trust environments, people often rely on these signals to assess others' credibility \cite{10.1145/22145.22178}. For example, when reading news articles, we must rely on cues such as author credibility and presentation quality rather than directly verifying the facts. GenAI systems can simulate these mental cues by constructing convincing, well‐structured arguments that lead users to mistake the AI's output for genuine responses, with cognitive depth and accuracy. \cite{wojtowicz2025underminingmentalproofai, kahneman2011thinking}

Similarly, \citet{harland2024aiapologycriticalreview} analyzed how AI writes apologies and divided important patterns (e.g., interaction, offence, recipient, and offender). They showed how these elements map onto the structure of effective apologies in human–machine contexts. For example, a customer-service bot that mishandles a transfer may explicitly note the mistake, recognize the user’s inconvenience, apologize, attribute the failure to a system issue, and commit to an immediate fix and prevention. This example illustrates how AI can convincingly simulate the cues of genuine remorse. \cite{harland2024aiapologycriticalreview, wojtowicz2025underminingmentalproofai} 

Moreover, by amplifying these simulated cues: consistent tone, structure, and ensuring statements, GenAI can trigger a false sense of rapport or expertise. Users subconsciously attribute a level of "understanding" or "intent" to the AI, leading them to accept its suggestions with little scrutiny. \cite{apologies}

\subsection{Overreliance on AI}
Given the convincing cues that AI can externalize \cite{harland2024aiapologycriticalreview, wojtowicz2025underminingmentalproofai, apologies}, a consistent finding in human-AI interaction research is the tendency for users to rely too heavily on AI suggestions. In this context, overreliance refers to accepting AI-generated suggestions even when those suggestions are incorrect. This overreliance on AI often leads to more incorrect decisions than those made independently, without AI assistance \cite{10.5555/2343576.2343643, 10.5555/3061053.3061219}. This unbalanced trust in AI output highlights the potential risks of overreliance on GenAI. \cite{vasconcelos2023explanationsreduceoverrelianceai, Bu_inca_2021, Kim_2025, bansal2021doesexceedpartseffect}

\citet{Bu_inca_2021} proposed cognitive forcing functions, which are interface-level interventions designed to stimulate user analytical thinking. These include requiring users to choose between viewing the AI suggestion and the AI-generated response confidence levels, among other measures. These efforts reduced overreliance. However, benefits often come at the cost of usability and user satisfaction.

\citet{vasconcelos2023explanationsreduceoverrelianceai} connected overreliance to a cost–benefit framework with a model that assesses the perceived benefits of using an AI suggestion in a cognitively costly task \cite{navon1979economy}. The authors demonstrated that people selectively engage in AI explanations, depending on the perceived difficulty of the task and the potential benefits that AI offers. In the context of our research, this framework provides reasons for achieving balanced experiences when utilizing AI to solve tasks. For example, when verifying the AI's recommendation is less cognitively costly than solving the task independently, users rely on it less.

\citet{Kim_2025} extended this discussion by showing that AI explanations and clickable sources influence user trust in AI. Although explanations increased trust in both correct and incorrect suggestions, sources had a more balanced effect, promoting appropriate reliance on correct suggestions and discouraging acceptance of incorrect ones. Clickable references embed links or buttons directly within the information source, allowing users to select them to reveal the original supporting material. These links encourage users to verify each assertion and trace its origin. Early studies find that providing access to source material can nudge users toward more critical evaluation \cite{Kim_2025, Tang2023}. However, users should remain cautious about fabricated sources, as shown by a study by \citet{hallucinations}. Relatedly, \citet{10.1145/3630106.3658941} found that when AI output contains uncertainty, for example, "\textit{I'm not sure, but\ldots}", users are less likely to overrely on incorrect responses, though the effect depends on how the uncertainty is phrased. This suggests that the form of AI assistance, not just its presence, shapes the extent to which users adopt it. 

\citet{vasconcelos2023explanationsreduceoverrelianceai} show that explanations, also known as Explainable AI or XAI, can prevent users from blindly accepting wrong suggestions, but only if those explanations make it easier or more worthwhile to check the AI’s advice than trust it. Explainable AI surfaces the model's underlying reasoning through logical rules or visual path highlights, helping users judge when to trust or question the AI's output. By formalizing explanation use within a cost–benefit framework, explanation clarity, and incentives, they show that easily digestible explanations reduce overreliance \cite{vasconcelos2023explanationsreduceoverrelianceai, Zhang_2020, ehsan2020humancenteredexplainableaireflective, 10.1145/3710946}.

\citet{toker2024roletaskcomplexityreducing} contributed to this discussion by demonstrating that increasing task complexity can reduce reliance on generative AI and minimize AI plagiarism. In their experiment, the results showed a decrease in AI-generated content as tasks progressed toward higher-order thinking. This supports the premise that designing cognitively demanding tasks can reduce reliance on AI output.

These findings emphasize the complexity of calibrating AI and trusting interfaces to avoid overreliance. Recent work has further formalized this challenge. \citet{10.1145/3630106.3658901} proposed a decision-theoretic framework that grounds reliance measurement in statistical decision theory, showing that human-AI teams frequently underperform the AI alone because users follow AI recommendations even when doing so leads to worse outcomes than relying on their own judgment. However, previous studies on overreliance have primarily examined multiple-choice decisions with binary, correct or incorrect answer options. Our work aims to extend this to open-ended tasks, where the effects can be more nuanced. Specifically, our multi-metric approach complements \citet{toker2024roletaskcomplexityreducing} and \citet{chen_more_2025} by providing quantitative similarity metrics that measure the degree to which AI suggestions shape written output, rather than inferring reliance from performance outcomes or self-reported measures alone. 

\subsection{Similarity Analysis Between Human and AI Text}
Prior work has examined the influence of AI in interactive writing contexts. CoAuthor \cite{Lee_2022}, and Wordcraft \cite{coenen2021wordcrafthumanaicollaborativeeditor} are two human-AI collaborative writing systems that study how users adopt or resist AI-generated suggestions during writing. CoAuthor provides a large-scale dataset of human writing sessions with GPT-3, revealing how writers incorporate and modify suggestions across different stages of writing. Wordcraft is a creative writing editor designed to support writers through a range of AI-assisted operations, including elaboration, rewriting, and continuation. Both systems primarily focus on behavioural patterns and user experience, leaving open the question of how to quantify the extent to which AI shapes the final text.

Research in AI-mediated communication (AMC) has similarly found measurable shifts in style and sentiment attributable to AI exposure. \citet{10.1145/3449091} identified a positivity bias in AI-generated communication, while \citet{10.1145/3544548.3581351} examined how sentence- versus message-level suggestions differ in adoption patterns, and \citet{10.1145/3701716.3717543} observed changes in text complexity and sentiment in social media posts following ChatGPT's public introduction.

Our study extends this line of inquiry. While CoAuthor and Wordcraft characterize how writers interact with AI behaviourally, and AMC studies document population-level linguistic shifts, we contribute a controlled experiment for quantifying adoption across multiple textual dimensions simultaneously. This approach allows us to decompose AI influence into distinct axes and test which dimensions are affected by different assistance formats.

\subsection{Impact on Cognitive Engagement}
Beyond trust and accuracy, researchers have also explored the cognitive implications of GenAI. In a survey with knowledge workers, \citet{10.1145/3706598.3713778} found that frequent use of GenAI tools was associated with a self-reported decrease in mental effort. Participants described a shift from "doing" to "overseeing" when using AI tools to solve day‐to‐day tasks. This shift may reduce the opportunity for cognitive development, particularly in tasks that require judgment and original thinking.

\citet{chen_more_2025} reported that users \emph{preferred} the automated AI assistance condition for perceived ease and less effort, describing this as a discrepancy between convenience preference and cognitive benefit.

In Bloom’s taxonomy framework, originally proposed by Benjamin Bloom in 1965 and widely used in educational settings, this shift corresponds to a lower level of engagement with higher-order cognitive processes. Rather than formulating their arguments (Analysis) or integrating diverse ideas into new perspectives (Synthesis), users often accept AI-produced text with limited alteration, thus undermining vital practice in these cognitive skills. \cite{bloom1956taxonomy, Hui07022025, singh2025protectinghumancognitionage}

On a physical level, \citet{kosmyna2025brainchatgptaccumulationcognitive} conducted a large-scale EEG study to investigate how external AI support influences neural indicators of cognitive engagement during essay writing. They found that participants who relied on LLMs exhibited significantly weaker neural connectivity than those who wrote without assistance. This pattern suggests that AI assistance immediately reduces cognitive effort and may reduce neural processes associated with analytical thinking and memory retrieval. 

Although it is not a central focus of our study, we also administered a TLX (cognitive load) questionnaire as a post-task self-reflection component of our main experiment. We elaborate on this in the Results and Discussion sections.

\section{Methods}
To investigate how users adopt AI suggestions into open-ended writing and how this aligns with users' own accounts of AI output usage, we conducted a within-participant, counterbalanced across tasks, user study, approved by the Simon Fraser University Research Ethics Board (30003047). For the main experiment, participants resolved two tasks based on a short reference text. We then prototyped and conducted a follow-up think-aloud evaluation of a writing interface intervention. 

\subsection{Materials and Tasks}
In the main experiment, participants completed two open-ended tasks based on the short story "After Twenty Years" by O. Henry \cite{OHenry1906AfterTwentyYears}. We chose this text because of its concise form, which allows quick reading while simultaneously requiring cognitive engagement for interpretation. Its twist ending and well-rounded characters provide participants with many points for analysis while minimizing background knowledge variance. For the writing interface intervention, participants completed an open-ended writing activity in which they wrote three paragraphs on a topic of their choice.

\subsection{Participants and Recruitment}
We recruited participants on Prolific across phases. Participants reported fluency in English and at least secondary-level education. We compensated the participants at \pounds6.00/hr, with an additional \pounds1.00 bonus in some variations of the experiment. For the think-aloud interface intervention study, we recruited undergraduate students from various course backgrounds.

To achieve precise similarity metrics in the main experiment, we explicitly instructed participants not to use any external large language models. Instead, they should use the AI suggestion provided in the interface of our experiment, if available. We manually examined the participants who pasted content in the experiment and found a high likelihood that they were produced by external LLMs, given their completion time and pasted content compared to the final response \cite{russell2025peoplefrequentlyusechatgpt, article}. A recent study found that 33-46\% of crowd-workers rely on LLMs for text production tasks \cite{veselovsky2023artificialartificialartificialintelligence}. Although using external LLMs for this study could be informative for the topic, we focused on task completion, as it could affect our measurements. We also consider this to warrant further exploration in future experiments. 

\subsection{Pilot Experiment}
We conducted a small pilot experiment in two iterations. First, a brief, time-limited pilot with 5 participants was conducted to validate the setup; it revealed that the time constraints and phrasing produced short, surface-level responses, limiting the interpretability of the adoption and reflection data \cite{gergle2014experimental, allen2023hurry}. We then conducted an untimed iteration with 11 participants using revised prompts and incentives. The pilot data were used solely to inform the design change and were not treated as evidence for the main experiment.

Across the pilot iterations, the key changes were: (i) removing strict time limits, (ii) removing "brief" and "short" prompt phrasing that encourages minimal responses both to the task and self-reflection, (iii) adding a small bonus to incentivize richer reflections, and (iv) changing the experiment design from between-subjects to a counterbalanced within-subjects design to reduce sensitivity to individual differences in writing ability and baseline similarity. These changes were motivated by our goal of comparing observed adoption with perceived adoption, which requires sufficiently rich responses and reflections.

\subsection{Main Experiment}
We recruited a total of 54 participants on Prolific. After data collection, we excluded 7 participants who used external tools, resulting in 47 valid participants. Each user completed two tasks: an analysis task and a synthesis task. The tasks were designed to be cognitively engaging to reduce trivial responses and better support conclusions about adoption behaviour \cite{toker2024roletaskcomplexityreducing}.

\paragraph{Assistance Format}
The study included two tasks: one analytical and one creative. In assisted trials, the format of AI assistance depended on the type of task. For the analytical task, participants received a fully direct AI response that addressed the prompt. For the creative task, participants received a scaffolded suggestion (e.g. key points and proposed structure). This asymmetric assistance design allowed us to compare the adoption pattern between two common GenAI support models. 

Before the study, the research team reviewed AI suggestions for both tasks to ensure factual accuracy and quality. No misleading or incorrect content was included, ensuring that observed adoption effects reflect natural exposure to real-scenario suggestions rather than a manipulation of suggestion quality or correctness.

\begin{figure}[ht]
  \centering
  \includegraphics[width=1\linewidth]{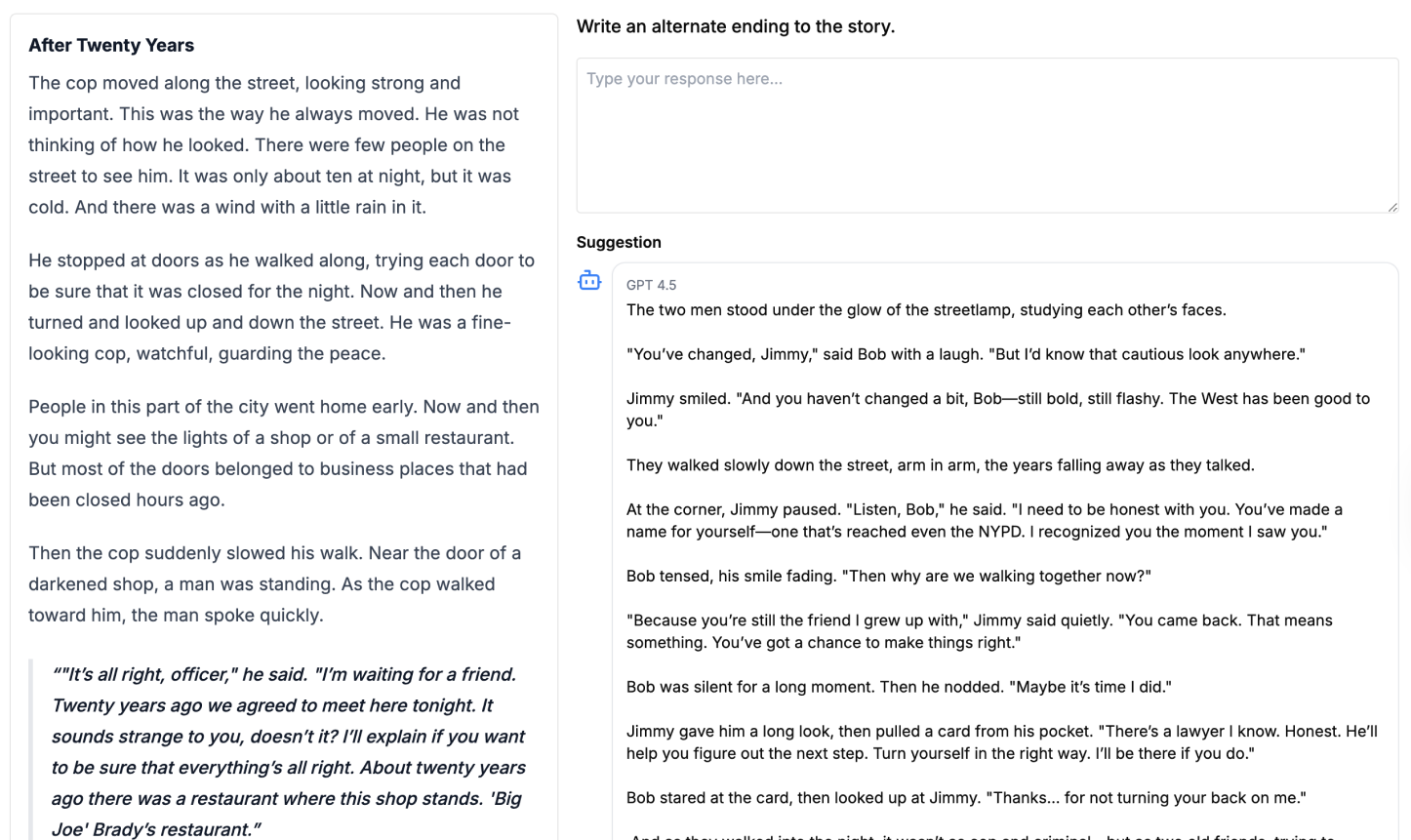}
  \caption{Experiment interface showing: (a) reading of the reference text, (b) presentation of AI suggestions, and (c) participant response entry field. These screenshots illustrate the environment in which participants completed both tasks.}
  \label{fig:experiment-setup}
\end{figure}

To illustrate the experimental setup and task interface, we include a screenshot in Figure~\ref{fig:experiment-setup}.

The specific task prompts and the AI suggestion used in both conditions are provided in Appendix~\ref{appendix:tasks}.

\paragraph{Counterbalancing and Assignment} 
To mitigate task order effects, task order and AI assistance were counterbalanced (Latin-square style). Participants faced an AI suggestion on at least one task, with the order of exposure varying between participants. This counterbalancing enables within-participant comparisons across tasks while allowing between-subject variation in order and task pairing.

\paragraph{Procedure}
The participants first read the story in the interface and then completed two tasks in sequence. The interface presented a prompt and a text editor, along with the AI suggestion panel for the assisted trials. Participants were not informed of the study's specific hypothesis. After each task, participants completed a brief post-task reflection survey, including the NASA-TLX questionnaire \cite{hart1988development}, which measures self-reported cognitive effort. In assisted trials, participants also completed open-ended reflection questions, from which the perceived level of AI suggestion use was derived through thematic coding.

\paragraph{Counterfactual baseline}
To compare AI-present and AI-absent writing using the same adoption metrics, we score every response against a fixed, task-specific AI suggestion text. In AI-present trials, this is the suggestion shown to the participant. In no-AI trials, participants don't see a suggestion; however, we still compute similarity between their response and the held-out task AI suggestion shown to other participants for the same prompt. We refer to these scores as a counterfactual baseline. For the creative task, the AI suggestion took the form of scaffold bullet points rather than a full direct answer. Accordingly, the counterfactual baseline for that task was computed against the bullet points, not against a hypothetical full-text response. Similarity scores across conditions should therefore be interpreted relative to each condition's baseline.

\paragraph{Similarity Metrics}
For each participant response $A$, we quantify similarity to the fixed AI suggestion of the task $B$. In trials without an AI suggestion, we still compare $A$ to the same task-level suggestion as a counterfactual baseline, allowing direct AI vs.\ no-AI comparisons. We report four complementary measures \cite{qurashi2020document}:
\begin{enumerate}
\item \textbf{Jaccard similarity (lexical overlap)} We compute the Jaccard coefficient \cite{jaccard1901} on sets of word tokens extracted with the regex pattern \verb|[A-Za-z']+|. The Jaccard similarity is then computed as the number of unique tokens shared by both texts divided by the number of unique tokens appearing in either text. Values closer to 0 indicate smaller lexical overlap, while values closer to 1 indicate greater lexical overlap.
\item \textbf{POS TF-ISF cosine (structural alignment)} Following \citet{vani2015investigating}, we perform part-of-speech tagging (POS-tagging), form lemma+class terms and compute the cosine similarity between the vectors. Values closer to 0 indicate weaker structure alignment, while values closer to 1 indicate stronger alignment.
\item \textbf{SBERT cosine similarity (semantic alignment)} We encode each text with Sentence-BERT \cite{reimers2019sentencebertsentenceembeddingsusing} using \texttt{roBERTa}\footnote{\url{https://huggingface.co/sentence-transformers/roberta-base-nli-mean-tokens}} and compute cosine similarity between the embeddings. Values closer to -1 indicate weaker semantic alignment, while values closer to 1 indicate stronger alignment. 
\item \textbf{Aspect sentiment match (sentiment alignment)} In line with the aspect-level sentiment consistency metric proposed by \citet{zhao_als-mrs_2022}, we split each text into sentences and compute the TextBlob polarity for each one \cite{loria2026textblob, abiola2023sentiment}. We then compare the aspect sentiment labels (positive/negative/neutral) between the participant response and the task suggestion. Values closer to 0 indicate weaker sentiment alignment, while values closer to 1 indicate stronger alignment.
\end{enumerate}

Together, these metrics capture adoption at the lexical, structural, semantic, and sentiment levels. We note that individual similarity values are expected to be high even in no-AI trials because both the participant response and the AI suggestion address the same short story. This expected baseline is precisely why relying on a single similarity metric is insufficient.

\paragraph{Workload (cognitive engagement).}
We collected a NASA-TLX-style workload measure after tasks as an exploratory context. We report these and interpret null/non-robust effects carefully, consistent with our goal of focusing on AI output adoption rather than cognitive outcomes.

\paragraph{Statistical hypothesis tests}
Across the analysis, we report condition means and standard deviations ($M$, $SD$), mean differences ($\Delta$), and effect sizes ($d_z$ for paired and Cohen's $d$ for independent-group designs). We conduct two-sided hypothesis tests at a significance level $\alpha = .05$. The null hypothesis for each metric is that AI presence does not change the outcome relative to no-AI. Conversely, the alternative hypothesis is that AI assistance changes the outcome.

Our statistical tests are exploratory and hypothesis-generating. We report mean differences and effect sizes to characterize the magnitude of observed patterns, and we treat $p$-values as descriptive indicators of strength rather than definitive evidence. This is important because we evaluate multiple related outcomes that may be correlated. Accordingly, we interpret the $p$-values cautiously (Tables~\ref{tab:anova_task_ai}, \ref{tab:similarity_task_ai}, \ref{tab:time_task_ai}, and \ref{tab:tlx_task_ai}). While future work could pre-register a primary adoption metric, we treat the set of metrics as exploratory and complementary.

\paragraph{Qualitative analysis.}
We analyzed post-task reflections from AI-present trials in the main experiment to examine what participants reported adopting or rejecting (ideas, structure, phrasing) and how they justify those choices. The first author conducted thematic coding without a pre-defined codebook, with categories generated inductively across multiple passes over the reflections. The second author reviewed the final results. The relatively small set of observations allowed for thorough analysis. Given the exploratory nature of the study and the relatively small sample, formal inter-rater reliability was not computed. We interpret the qualitative findings as descriptive.

\subsection{Think-Aloud Study of the Implemented Interface Intervention}
To evaluate whether an interface intervention can shape how users interpret and incorporate AI suggestions during writing, we conducted a follow-up think-aloud study using the implemented interface.

\begin{figure}
    \centering
    \includegraphics[width=1\linewidth]{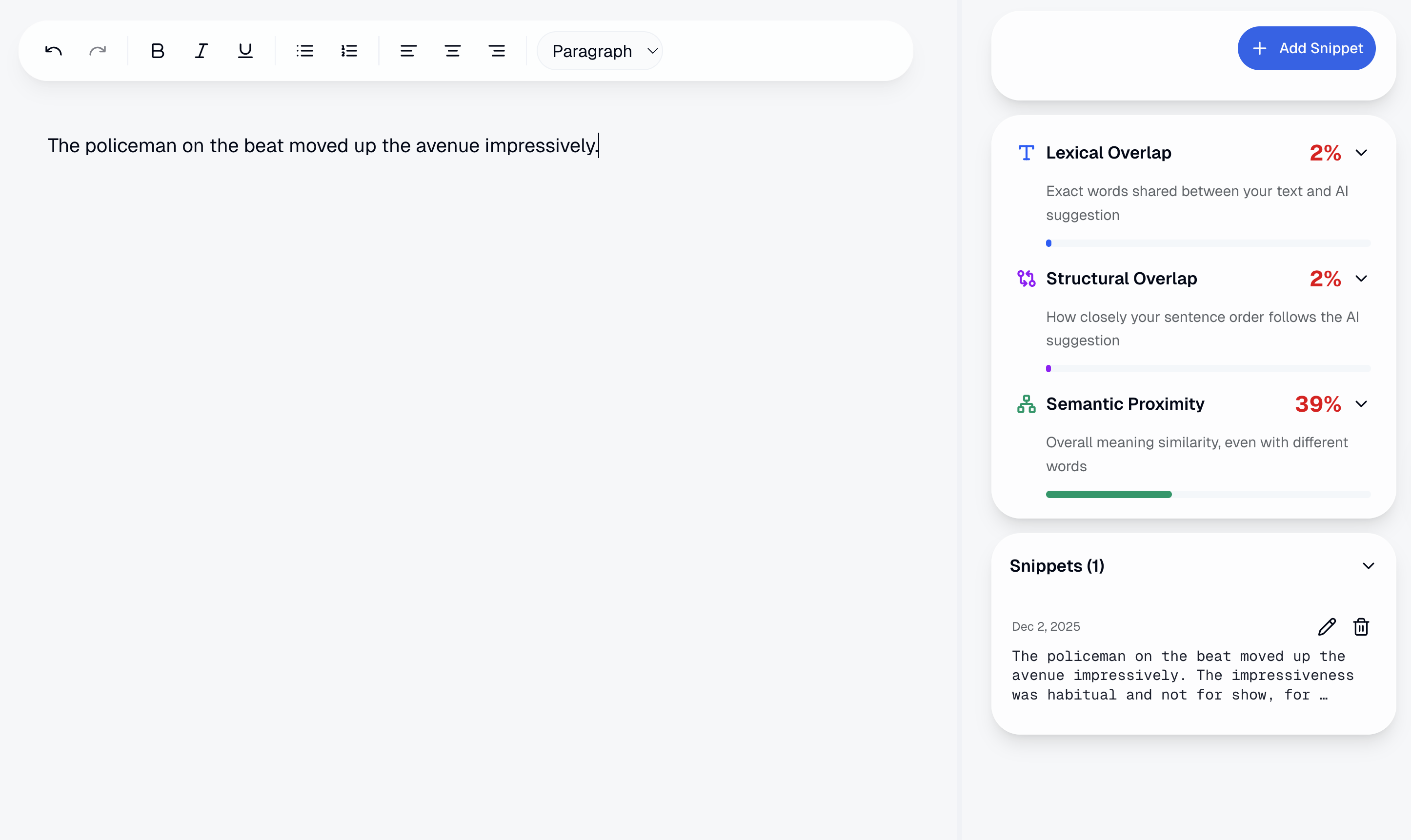}
    \caption{Participants draft text in the main editor (left) while a side panel (right) provides feedback on similarity to an AI suggestion and saved snippets of AI outputs for reference and review.}
    \label{fig:reflective-writing-interface}
\end{figure}

\paragraph{Interface.}
We prototyped a web-based editor in which participants could paste any AI output they used as reference “snippets.” The participants completed the task entirely within the editor. As they wrote, the sidebar indicators updated in real time to show how closely the draft aligned with the added snippets, making the adoption more noticeable without constraining whether participants accepted, edited or ignored the suggestions. We include a representative screenshot in Figure~\ref{fig:reflective-writing-interface}. 

\paragraph{Procedure.}
Participants completed an open-ended writing activity by composing three paragraphs on a topic of their choice. For this task, they could use any LLM however they wished (e.g., brainstorming, drafting, or proofreading) and were asked to paste any AI output they referenced into the application as snippets. We specifically permitted the use of external LLMs for this task, as participants were able to record the exact suggestions they consulted, allowing us to compare their writing directly with those outputs and compute similarity measures accurately.

Before starting, we provided a quick overview of the editor, the snippet panel, and the real-time alignment feedback. We recorded the screen and audio and collected written debrief responses. After the task, participants took part in a debrief interview focused on (i) how they decided to adopt or reject aspects of the AI output (e.g., phrasing, structure, ideas) and (ii) which interface elements, if any, made AI influence more noticeable or easier to reflect on.

\paragraph{Analysis.}
The first author open-coded the transcripts, focusing on two phenomena identified in advance: moments where participants made a clear decision to adopt, edit, or reject part of an AI suggestion, and moments where the sidebar metrics influenced participants. Given the small sample ($n = 4$), we treat the analysis as illustrative and use direct quotes to ground each theme.

\section{Results}
Across 47 participants in a counterbalanced within-subject design, \textbf{exposure to an AI suggestion increased alignment between participants' writing and the AI output on multiple adoption measures}. Overall, AI-present trials showed higher lexical, structural, and aspect-level sentiment similarity than no-AI trials, whereas semantic similarity increased but did not reach statistical significance. These effects were driven primarily by the analytical task, where participants received a fully formed direct answer from the AI. In the creative task, where participants received a scaffolded suggestion, adoption effects were smaller, and only sentiment alignment increased significantly.

Self-reflections that reported using the suggestion often described it as helpful for direction and framing rather than as verbatim copying. \textbf{These results motivate measuring AI influence in writing as multi-dimensional adoption rather than binary copying}, and they set up our interface intervention aimed at making this influence more visible.

\subsection{Main Experiment}
\paragraph{Similarity Metrics.}
To quantify the adoption of AI output, we measured the similarity between each participant’s response and the AI's task suggestion using four complementary metrics (lexical, structural, semantic, and sentiment alignment). For no-AI trials, we compared responses to the same task suggestion as a counterfactual baseline.

Overall (Table~\ref{tab:anova_task_ai}), AI assistance increased similarity on lexical overlap (Jaccard: $M=0.108$ vs.\ $0.093$, $\Delta=0.015$, $p=.050$, $d_z=0.294$), structural similarity (POS TF-ISF cosine: $M=0.042$ vs.\ $0.025$, $\Delta=0.017$, $p=.029$, $d_z=0.329$), and aspect sentiment match ($M=0.092$ vs.\ $0.040$, $\Delta=0.052$, $p=.004$, $d_z=0.444$). SBERT cosine was higher with AI ($M=0.676$ vs.\ $0.642$, $\Delta=0.034$) but not statistically significant ($p=.086$, $d_z=0.256$).

Comparisons within tasks (Table~\ref{tab:similarity_task_ai}) indicate that this pattern is primarily driven by the analytical task, where we provided participants with a direct-answer AI suggestion. In that task, AI assistance increased Jaccard similarity ($0.128$ vs.\ $0.104$, $\Delta=0.024$, $p=.003$, $d=0.900$), POS TF-ISF cosine ($0.061$ vs.\ $0.035$, $\Delta=0.026$, $p=.022$, $d=0.692$), and aspect sentiment match ($0.147$ vs.\ $0.070$, $\Delta=0.077$, $p=.002$, $d=0.945$), while SBERT cosine was not significantly different ($p=.257$). In the creative task, where we provided participants with a scaffold AI suggestion, only the aspect sentiment match increased significantly ($0.036$ vs.\ $0.011$, $\Delta=0.024$, $p=.021$, $d=0.695$); differences in Jaccard, POS TF-ISF cosine, and SBERT cosine were not significant ($ps\ge .129$). 

Together, these results indicate that \textbf{AI assistance increased alignment with the suggestion, with the strongest adoption effects in the direct response AI suggestion condition}.

\begin{table}
\caption{Overall AI vs. no-AI differences in similarity (paired-samples t-test; paired by participant; tasks counterbalanced; significant rows bolded and marked with *).}
\label{tab:anova_task_ai}
\begin{tabular}{lrrrrr}
\toprule
Metric & No-AI, M (SD)& AI, M (SD)& $\Delta$ & $p$ & Cohen's $d_z$ \\
\midrule
\textbf{Jaccard} & \textbf{0.093 (0.031)} & \textbf{0.108 (0.038)} & \textbf{0.015} & \textbf{0.050*} & \textbf{0.294} \\
\textbf{POS TF-ISF cosine} & \textbf{0.025 (0.024)} & \textbf{0.042 (0.044)} & \textbf{0.017} & \textbf{0.029*} & \textbf{0.329} \\
SBERT (RoBERTa) cosine & 0.642 (0.104) & 0.676 (0.082) & 0.034 & 0.086 & 0.256 \\
\textbf{Aspect sentiment match} & \textbf{0.040 (0.055)} & \textbf{0.092 (0.093)} & \textbf{0.052} & \textbf{0.004*} & \textbf{0.444} \\
\bottomrule
\end{tabular}
\end{table}

\begin{table}
\caption{Within-task AI vs. no-AI differences in similarity to the task suggestion (independent-samples t-test; significant rows bolded and marked with *).}
\label{tab:similarity_task_ai}
\begin{tabular}{llrrrrr}
\toprule
Task & Metric & No-AI, M (SD)& AI, M (SD)& $\Delta$& $p$ & Cohen's $d$ \\
\midrule
Analytical& \textbf{Jaccard}& \textbf{0.104 (0.023)} & \textbf{0.128 (0.029)} & \textbf{0.024} & \textbf{0.003*} & \textbf{0.900} \\
\textbf{} & \textbf{POS TF-ISF cosine} & \textbf{0.035 (0.026)} & \textbf{0.061 (0.046)} & \textbf{0.026} & \textbf{0.022*} & \textbf{0.692} \\
 & SBERT (RoBERTa) cosine & 0.670 (0.128) & 0.707 (0.091) & 0.037 & 0.257 & 0.335 \\
\textbf{} & \textbf{Sentiment match}& \textbf{0.070 (0.061)} & \textbf{0.147 (0.097)} & \textbf{0.077} & \textbf{0.002*} & \textbf{0.945} \\
Creative & Jaccard & 0.082 (0.034) & 0.087 (0.035) & 0.006 & 0.582 & 0.162 \\
 & POS TF-ISF cosine & 0.015 (0.016) & 0.021 (0.031) & 0.006 & 0.393 & 0.252 \\
 & SBERT (RoBERTa) cosine & 0.615 (0.065) & 0.643 (0.058) & 0.028 & 0.129 & 0.451 \\
\textbf{} & \textbf{Sentiment match}& \textbf{0.011 (0.028)} & \textbf{0.036 (0.041)} & \textbf{0.024} & \textbf{0.021*} & \textbf{0.695} \\
\bottomrule
\end{tabular}
\end{table}

\paragraph{Self-Reported Usage and Qualitative Analysis.}
After each AI-assisted trial, participants completed a brief post-task reflection survey that included two open-ended questions asking (i) whether or how they used the given AI suggestion and (ii) whether they would have completed the task differently if not displayed with a suggestion. Given that we had 47 participants, each of whom completed a task with and without assistance, we collected 47 instances of reflection in total. 

Overall, 37/47 (78.7\%) of AI-present responses reported using the AI suggestion: 22 from the creative task and 15 from the analytical task. When participants used the suggestion, they more often described it as beneficial than not (26/37, 70.3\%), whereas 10/47 (21.3\%) of all participants who were shown the AI suggestion explicitly reported that it was not helpful. 

In terms of how they perceived themselves to adopt the suggestion, 22/37 (59.5\%) responses referenced the suggestion that gave direction, compared to 10/37 (27.0\%) mentioned wording or paraphrasing. These patterns were similar in proportion across task types. For the creative task, 16/22 (72.7\%) responses were rated as helpful, whereas for the analytical task, 10/15 (66.7\%) were rated as helpful.

For the second question, in which participants were asked whether they would complete the task differently without an AI suggestion, 28/47 (59.6\%) AI-present tasks were tagged as indicating that they would answer differently. Of these, 15/47 (31.9\%) referenced that the idea/content would be different, and 11/47 (24.5\%) referenced that the wording/grammar would be different.

In summary, these provide a complementary lens on AI adoption, in which \textbf{most participants perceived themselves as using the suggestion} and \textbf{many anticipated that their responses would change without it}. This is directly connected to our study goals of measuring AI output adoption and perceived influence, offering a user-centred account of how suggestions shape writing even when they are not described as direct copying.

\paragraph{Cognitive Load and Time.}
In general, \textbf{we did not observe reliable differences between AI and non-AI conditions in total TLX scores or completion time for either task} (Table~\ref{tab:tlx_task_ai} and Table~\ref{tab:time_task_ai}). 

\begin{table}
\caption{Within-task AI vs. no-AI differences in NASA-TLX ratings (independent-samples t-test; significant rows bolded and marked with *).}
\label{tab:tlx_task_ai}
\begin{tabular}{llrrrrr}
\toprule
Task & Metric & No-AI, M (SD)& AI, M (SD)& $\Delta$ & $p$ & Cohen's $d$ \\
\midrule
Analytical & Mental demand & 5.043 (1.821) & 4.667 (1.685) & -0.377 & 0.465 & -0.215 \\
 & Physical demand & 2.435 (1.532) & 2.917 (1.692) & 0.482 & 0.312 & 0.298 \\
 & Rushed & 2.652 (1.301) & 2.250 (1.152) & -0.402 & 0.267 & -0.328 \\
 & Accomplishment & 3.478 (1.563) & 3.458 (2.043) & -0.020 & 0.970 & -0.011 \\
 & Effort & 5.087 (1.564) & 4.792 (1.474) & -0.295 & 0.509 & -0.194 \\
 & Insecurity & 2.957 (1.364) & 2.208 (1.444) & -0.748 & 0.075 & -0.532 \\
 & TLX total & 3.609 (0.799) & 3.382 (0.888) & -0.227 & 0.363 & -0.268 \\
Creative & Mental demand & 5.250 (1.539) & 5.609 (1.438) & 0.359 & 0.414 & 0.241 \\
 & Physical demand & 3.458 (2.064) & 2.783 (1.536) & -0.676 & 0.211 & -0.370 \\
 & Rushed & 2.292 (1.459) & 3.130 (1.740) & 0.839 & 0.080 & 0.523 \\
 & Accomplishment & 3.500 (1.818) & 3.522 (1.880) & 0.022 & 0.968 & 0.012 \\
\textbf{} & \textbf{Effort} & \textbf{4.917 (1.472)} & \textbf{5.739 (1.176)} & \textbf{0.822} & \textbf{0.040*} & \textbf{0.616} \\
 & Insecurity & 2.500 (1.532) & 2.913 (1.474) & 0.413 & 0.352 & 0.275 \\
 & TLX total & 3.653 (0.965) & 3.949 (0.748) & 0.296 & 0.247 & 0.342 \\
\bottomrule
\end{tabular}
\end{table}

\begin{table}
\caption{Within-task AI vs. no-AI differences in completion time in minutes (independent-samples t-test; significant rows bolded and marked with *).}
\label{tab:time_task_ai}
\begin{tabular}{llrrrrr}
\toprule
Task & Metric & No-AI M (SD) & AI M (SD) & $\Delta$ & $p$ & Cohen's $d$ \\
\midrule
Analytical & Completion time (min) & 15.204 (9.785) & 18.246 (10.321) & 3.042 & 0.306 & 0.302 \\
Creative & Completion time (min) & 16.250 (8.905) & 20.875 (13.861) & 4.625 & 0.178 & 0.399 \\
\bottomrule
\end{tabular}
\end{table}

Most comparisons did not reach statistical significance. In the analytical task, AI assistance was associated with slightly lower mental demand, rushedness, effort, and insecurity. In the creative task, AI assistance was associated with slightly higher effort (the only statistically significant comparison), mental demand and rushedness, and a higher TLX total, although these differences were not statistically significant (all p > .05). The completion time also did not differ significantly between the AI and non-AI conditions for either task (Table \ref{tab:time_task_ai}). 

These results suggest that AI suggestions did not reliably reduce perceived workload or time-on-task, and any potential effects are likely small or context-dependent. Larger-scale studies would be valuable for estimating these effects more precisely and for testing whether they reliably emerge for particular task types and assistance formats.

\subsection{Think Aloud Experiment on the Proposed Writing Interface Intervention}
To evaluate whether real-time similarity feedback changes how users perceive and incorporate AI suggestions during writing, we conducted a think-aloud study with four university students using our reflective writing editor intervention. The interface was directly informed by the main experiment findings: given that participants showed measurable AI adoption across multiple dimensions yet often described their use as limited to ideas or direction, we designed feedback that makes these dimensions visible in real time, targeting the gap between perceived and actual adoption.

\paragraph{The metrics shaped writing strategies and calibrated reliance}
\textbf{Participants frequently described the sidebar metrics as a calibration aid that shaped writing choices} (e.g., paraphrasing, restructuring, or deciding when to stop consulting the AI or to ask for a follow-up). The value of the interface was often framed as not preventing or impeding work, but as helping participants decide how much to use AI. 

Participant 3 directly noted that "[It] was a very helpful tool in ensuring I was mindful of my AI usage," while Participant 4 emphasized its value as a boundary: “as a reluctant AI user, it’s helping me… in that kind of ethical way of… being aware of the impact.” 

Participant 1’s reflection was slightly different: they described using AI mainly when they “don’t know the wording,” and the interface helped them feel reassured that reliance was not as high as expected: “I definitely thought… the structure would have been a little bit higher.”

Interestingly, \textbf{calibration was not always about minimizing reliance}. During the task, Participant 2 reflected: "I’m using a bit of the suggestion… but I could rely more on it from what I’m seeing on the panel.” This highlights an important nuance: depending on users' goals, the same feedback can serve as a calibrated signal rather than uniformly discourage AI use.

\paragraph{The interface made AI influence identifiable during writing}
Most participants used the bars as a real-time mirror of AI use. Most reported observing changes in the indicators while copying or paraphrasing AI-generated text. As participant 4 noted, the bar "jumped up right when I copied and pasted something in there. So I was like, OK, that was helpful because like I definitely did that." Participant 3 described trying to make the bars "go lower, like down, down, down," treating the indicator as a live constraint that supported mindful writing. Participant 2 said they "always use AI detectors at the end of my papers to make sure that it's actually consistent with what I wrote. So having this going on as I'm writing the paper would save me time in the end."

These findings demonstrate that students can reflect on and adjust their adoption in real time, rather than treating it as a policing mechanism. By showing that participants used the bars to monitor, calibrate, and strategically adjust their engagement with AI, this study illustrates that the tool can be helpful as reflective writing support.

\section{Discussion}
Our main experiment results suggest that seeing generative AI output can measurably shape open-ended writing, even when participants do not copy verbatim. Across analytical and creative tasks, exposure to an AI suggestion increased alignment between participants’ responses and the AI output in lexical, structural, and sentiment overlap. This effect was strongest when participants were shown a direct-answer output rather than a scaffolded suggestion. Together, these findings support our core motivation: AI influence on writing is not binary. It often appears as a subtle adoption of wording, organization, and even affect, and these measurements should be available for users' reflection, which informed our reflective writing interface intervention.

\paragraph{Adoption vs. Overreliance}
It is important to sharpen the discussion between adoption, which we measure, and overreliance, which carries normative weight. Our study quantified textual alignment between participant responses and AI suggestions. It does not assess whether any given instance of adoption affected performance or correctness outcomes. Adoption becomes a concern, though not exclusively, (i) when the user is unaware of the extent to which AI has shaped their writing, (ii) when adoption goes against the purpose of the task, for example, in educational assessments designed to elicit original reasoning, and (iii) when the resulting text misrepresents the author's perspective or views. Determining whether and when observed adoption entails overreliance and effects on the mentioned concerns requires additional outcome measures, such as writing quality assessment, learning outcomes, task alignment, and correctness, which we identify as important directions for future work. In the current study, we use the term adoption to describe what we measure and reserve the term overreliance as a normative label for contexts in which this is demonstrated to be harmful.

\paragraph{AI output adoption is multi-dimensional}
We argue that AI influence in writing is not monolithic, and our findings reinforce that a single metric may miss meaningful forms of adoption. Some forms are not obvious to the naked eye, such as sentiment alignment. At the same time, semantic similarity increased but did not reliably reach significance, suggesting that the metrics we used capture only part of the influence landscape. In practice, there may be other dimensions of similarity that matter for writing that are not fully represented by standard semantic measures. This motivates treating AI influence as a multi-dimensional phenomenon rather than reducing it to a single indicator.

The clearest similarity lever in our study was the assistance format. Direct-answer suggestions in the analytical task produced large and consistent adoption effects, while scaffolded suggestions yielded smaller effects. This has direct design implications: reducing unintended or excessive adoption on written tasks may be less about adding warnings and more about changing the form of help. Scaffolds may better preserve degrees of freedom, based on the limited measurements from our study. However, scaffolds primarily address surface-level expression, whereas overreliance on AI-generated ideas and arguments can occur regardless of assistance formats, since users may still adopt the content of the AI suggestion. More broadly, our results reinforce that open-ended writing is a distinct interaction setting from binary decision-making and that interventions should be guided by the purpose of writing; for example, this may apply differently across learning and professional environments.

\paragraph{Reflective metrics}
Although not directly measured in our main experiment, our interface intervention addresses a potential gap between participants’ self-reported use of suggestions and their actual alignment with AI output. Reflective similarity feedback may function as a lightweight cognitive forcing mechanism: it introduces a moment of reflection without blocking progress. This shift emphasizes the opportunity to move away from policing toward agency—supporting user awareness rather than external enforcement. At the same time, reflective tools should be designed carefully so they do not become compliance instruments, or implicitly define “good writing” as merely “low similarity.” The goal is not to prohibit reuse, but to make adoption choices more visible and reflective.

\paragraph{Null and mixed results on workload}
We did not observe reliable reductions in perceived workload or time-on-task. This is important evidence that AI assistance may not uniformly reduce writing effort. One interpretation is that AI shifts where effort is spent rather than reducing it, consistent with prior work \citet{10.1145/3706598.3713778}. Users may expend less effort generating raw content but more effort evaluating and integrating suggestions. In open-ended tasks, especially, deciding whether to follow or resist a well-formed suggestion can be cognitively demanding.

\paragraph{Broader implications for adoption and authorship}
Our work extends the discussion of AI influence on writing beyond plagiarism and correctness. We treat it as a multidimensional issue of similarity and reflection: not all reuse is harmful, but users should have visibility into how much the system shapes what they write. This reframes the ethical concern from solely external policy enforcement to a more balanced, user-centric model of reflective empowerment. One that can support education and accountable authorship. The stakes of adoption differ across contexts: in educational settings, AI influence may undermine assessments, while in professional contexts, it raises questions about accountability. Future work should explore the real-world consequences of AI adoption in specific contexts.

\paragraph{Limitations and future work}
Our study is scoped to the tasks and settings we evaluated, and replication across genres and contexts would strengthen generalizability. One limitation of the current design is that task type and assistance format co-vary: the analytical task used a direct-answer suggestion while the creative task used a scaffolded suggestion. This means that we cannot fully determine whether the stronger adoption effects in the analytical condition reflect the format of assistance, the nature of the task, or an interaction between the two. Future work that focuses specifically on measuring these interactions should independently vary task type and assistance format.

Another consideration is the ecological validity of our sample. Participants were Prolific crowd-workers completing a low-stakes short-story analysis task, which may not generalize to high-stakes or professional settings. Additionally, prior familiarity with the O. Henry reference text was not pre-screened. However, since both AI and no-AI conditions use the same reference text in a counterbalanced design, any differential familiarity is unlikely to bias AI versus no-AI comparisons.

The follow-up interface evaluation should be understood as a design provocation, an initial exploration of where real-time similarity feedback can support reflective writing. With only four participants, we cannot make strong claims about usability or effect. The open-ended nature of the topics in the interface study reflects a naturalistic AI-assisted writing condition and allows us to compute similarity directly against the AI output each participant actually consulted. However, this also introduces variability that limits systematic cross-participant comparison. Future work should extend this with a controlled design setting.

\section{Conclusion}
Generative AI output can shape open-ended writing in ways that extend beyond verbatim copying. In this paper, we examined how exposure to AI suggestions influences users’ written responses to analysis and synthesis tasks, leading to a multidimensional alignment between a participant’s text and an AI suggestion. In our main experiment, AI-assisted trials showed higher alignment than no-AI trials on lexical overlap, structural similarity, and aspect-level sentiment match, whereas semantic similarity showed a trend upward but was not consistently significant. The strongest effects appeared when participants received a fully formed direct-answer AI suggestion rather than a scaffolded outline. 

To complement these similarity measures, we designed an interface intervention that surfaces real-time similarity feedback during writing, aiming to make AI influence more visible. In a follow-up think-aloud study, participants treated the live indicators as a calibration aid, using them to monitor reuse, decide when to paraphrase, and reflect on how much they wanted AI to shape their text. This reflective feedback can support more conscious engagement without positioning the tool as a policing mechanism. 

Together, our findings argue for studying AI influence on writing as a subtle, multi-axis form of adoption, and for interventions that support author agency through visibility rather than restriction. Future work should test across genres and real-world settings, scale the experiment, and examine how more diverse individuals moderate influence.

\section*{Generative AI Usage Statement}
We used ChatGPT (OpenAI; GPT-5.2 Thinking) for grammar/style proofreading and to assist with drafting code to generate the paper’s figures from our existing data. All results were reviewed and edited by the authors, who remain fully responsible for the manuscript’s originality and integrity.

\bibliographystyle{ACM-Reference-Format}
\bibliography{references}

@String{Computing = "Computing" }

@String{Springer = "Springer-Verlag" }

@misc{veselovsky2023artificialartificialartificialintelligence,
      title={Artificial Artificial Artificial Intelligence: Crowd Workers Widely Use Large Language Models for Text Production Tasks}, 
      author={Veniamin Veselovsky and Manoel Horta Ribeiro and Robert West},
      year={2023},
      eprint={2306.07899},
      archivePrefix={arXiv},
      primaryClass={cs.CL},
      url={https://arxiv.org/abs/2306.07899}, 
}

@inproceedings{bansal2021doesexceedpartseffect,
author = {Bansal, Gagan and Wu, Tongshuang and Zhou, Joyce and Fok, Raymond and Nushi, Besmira and Kamar, Ece and Ribeiro, Marco Tulio and Weld, Daniel},
title = {Does the Whole Exceed its Parts? The Effect of AI Explanations on Complementary Team Performance},
year = {2021},
isbn = {9781450380966},
publisher = {Association for Computing Machinery},
address = {New York, NY, USA},
url = {https://doi.org/10.1145/3411764.3445717},
doi = {10.1145/3411764.3445717},
booktitle = {Proceedings of the 2021 CHI Conference on Human Factors in Computing Systems},
articleno = {81},
numpages = {16},
location = {Yokohama, Japan},
series = {CHI '21}
}

@article{vasconcelos2023explanationsreduceoverrelianceai,
  title={Explanations can reduce overreliance on ai systems during decision-making},
  author={Vasconcelos, Helena and J{\"o}rke, Matthew and Grunde-McLaughlin, Madeleine and Gerstenberg, Tobias and Bernstein, Michael S and Krishna, Ranjay},
  journal={Proceedings of the ACM on Human-Computer Interaction},
  volume={7},
  number={CSCW1},
  pages={1--38},
  year={2023},
  publisher={ACM New York, NY, USA}
}

@inproceedings{Kim_2025, series={CHI ’25},
   title={Fostering Appropriate Reliance on Large Language Models: The Role of Explanations, Sources, and Inconsistencies},
   url={http://dx.doi.org/10.1145/3706598.3714020},
   DOI={10.1145/3706598.3714020},
   booktitle={Proceedings of the 2025 CHI Conference on Human Factors in Computing Systems},
   publisher={ACM},
   author={Kim, Sunnie S. Y. and Vaughan, Jennifer Wortman and Liao, Q. Vera and Lombrozo, Tania and Russakovsky, Olga},
   year={2025},
   month=apr, pages={1–19},
   collection={CHI ’25} }

@article{article,
author = {Al-Rawas, Matheel and Qader, Omar and Othman, Nurul and Ismail, Noor and Mamat, Rosnani and Halim, Mohamad Syahrizal and Abdullah, Johari and Noorani, Tahir},
year = {2025},
month = {04},
pages = {},
title = {Identification of dental related ChatGPT generated abstracts by senior and young academicians versus artificial intelligence detectors and a similarity detector},
volume = {15},
journal = {Scientific Reports},
doi = {10.1038/s41598-025-95387-y}
}

@inproceedings{russell2025peoplefrequentlyusechatgpt,
  title={People who frequently use ChatGPT for writing tasks are accurate and robust detectors of AI-generated text},
  author={Russell, Jenna and Karpinska, Marzena and Iyyer, Mohit},
  booktitle={Proceedings of the 63rd Annual Meeting of the Association for Computational Linguistics (Volume 1: Long Papers)},
  pages={5342--5373},
  year={2025}
}

@inproceedings{10.1145/3706598.3713778,
author = {Lee, Hao-Ping (Hank) and Sarkar, Advait and Tankelevitch, Lev and Drosos, Ian and Rintel, Sean and Banks, Richard and Wilson, Nicholas},
title = {The Impact of Generative AI on Critical Thinking: Self-Reported Reductions in Cognitive Effort and Confidence Effects From a Survey of Knowledge Workers},
year = {2025},
isbn = {9798400713941},
publisher = {Association for Computing Machinery},
address = {New York, NY, USA},
url = {https://doi.org/10.1145/3706598.3713778},
doi = {10.1145/3706598.3713778},
booktitle = {Proceedings of the 2025 CHI Conference on Human Factors in Computing Systems},
articleno = {1121},
numpages = {22},
location = {
},
series = {CHI '25}
}

@misc{toker2024roletaskcomplexityreducing,
      title={The Role of Task Complexity in Reducing AI Plagiarism: A Study of Generative AI Tools}, 
      author={Sacip Toker and Mahir Akgun},
      year={2024},
      eprint={2412.13412},
      archivePrefix={arXiv},
      primaryClass={cs.HC},
      url={https://arxiv.org/abs/2412.13412}, 
}

@article{Bu_inca_2021,
   title={To Trust or to Think: Cognitive Forcing Functions Can Reduce Overreliance on AI in AI-assisted Decision-making},
   volume={5},
   ISSN={2573-0142},
   url={http://dx.doi.org/10.1145/3449287},
   DOI={10.1145/3449287},
   number={CSCW1},
   journal={Proceedings of the ACM on Human-Computer Interaction},
   publisher={Association for Computing Machinery (ACM)},
   author={Buçinca, Zana and Malaya, Maja Barbara and Gajos, Krzysztof Z.},
   year={2021},
   month=apr, pages={1–21} }

@inproceedings{wojtowicz2025underminingmentalproofai,
author = {Wojtowicz, Zachary and DeDeo, Simon},
title = {Undermining mental proof: how AI can make cooperation harder by making thinking easier},
year = {2025},
isbn = {978-1-57735-897-8},
publisher = {AAAI Press},
url = {https://doi.org/10.1609/aaai.v39i2.32151},
doi = {10.1609/aaai.v39i2.32151},
booktitle = {Proceedings of the Thirty-Ninth AAAI Conference on Artificial Intelligence and Thirty-Seventh Conference on Innovative Applications of Artificial Intelligence and Fifteenth Symposium on Educational Advances in Artificial Intelligence},
articleno = {178},
numpages = {9},
series = {AAAI'25/IAAI'25/EAAI'25}
}

@article{AIwriting,
  title={The AI Revolution in Education: Will AI Replace or Assist Teachers in Higher Education?},
  author={Cecilia Ka Yuk Chan and Louisa H.Y. Tsi},
  journal={ArXiv},
  year={2023},
  volume={abs/2305.01185},
  url={https://api.semanticscholar.org/CorpusID:258436716}
}

@misc{kosmyna2025brainchatgptaccumulationcognitive,
      title={Your Brain on ChatGPT: Accumulation of Cognitive Debt when Using an AI Assistant for Essay Writing Task}, 
      author={Nataliya Kosmyna and Eugene Hauptmann and Ye Tong Yuan and Jessica Situ and Xian-Hao Liao and Ashly Vivian Beresnitzky and Iris Braunstein and Pattie Maes},
      year={2025},
      eprint={2506.08872},
      archivePrefix={arXiv},
      primaryClass={cs.AI},
      url={https://arxiv.org/abs/2506.08872}, 
}

@incollection{OHenry1906AfterTwentyYears,
  author       = {O. Henry},
  title        = {After Twenty Years},
  booktitle    = {The Four Million},
  publisher    = {McClure, Phillips \& Co.},
  address      = {New York},
  year         = {1906},
  note         = {Originally published in 1906; short story},
}

@incollection{hart1988development,
  title={Development of NASA-TLX (Task Load Index): Results of empirical and theoretical research},
  author={Hart, Sandra G and Staveland, Lowell E},
  booktitle={Advances in psychology},
  volume={52},
  pages={139--183},
  year={1988},
  publisher={Elsevier}
}

@book{bloom1956taxonomy,
  title={Taxonomy of educational objectives: The classification of educational goals. Handbook 1: Cognitive domain},
  author={Bloom, Benjamin S and Engelhart, Max D and Furst, Edward J and Hill, Walker H and Krathwohl, David R and others},
  year={1956},
  publisher={Longman New York}
}

@article{abiola2023sentiment,
  title={Sentiment analysis of COVID-19 tweets from selected hashtags in Nigeria using VADER and Text Blob analyser},
  author={Abiola, Odeyinka and Abayomi-Alli, Adebayo and Tale, Oluwasefunmi Arogundade and Misra, Sanjay and Abayomi-Alli, Olusola},
  journal={Journal of Electrical Systems and Information Technology},
  volume={10},
  number={1},
  pages={5},
  year={2023},
  publisher={Springer}
}

@article{harland2024aiapologycriticalreview,
  title={AI apology: a critical review of apology in AI systems},
  author={Harland, Hadassah and Dazeley, Richard and Senaratne, Hashini and Vamplew, Peter and Cruz, Francisco and Nakisa, Bahareh},
  journal={Artificial Intelligence Review},
  volume={58},
  number={12},
  pages={369},
  year={2025},
  publisher={Springer}
}

@article{understandingandsharingintentions,
author = {Tomasello, Michael and Carpenter, Malinda and Call, Josep and Behne, Tanya and Moll, Henrike},
year = {2005},
month = {11},
pages = {675-735},
title = {Understanding and Sharing Intentions: The Origins of Cultural Cognition},
volume = {28},
journal = {Behavioral and Brain Sciences},
doi = {10.1017/S0140525X05000129}
}

@inproceedings{10.1145/22145.22178,
author = {Goldwasser, S and Micali, S and Rackoff, C},
title = {The knowledge complexity of interactive proof-systems},
year = {1985},
isbn = {0897911512},
publisher = {Association for Computing Machinery},
address = {New York, NY, USA},
url = {https://doi.org/10.1145/22145.22178},
doi = {10.1145/22145.22178},
booktitle = {Proceedings of the Seventeenth Annual ACM Symposium on Theory of Computing},
pages = {291–304},
numpages = {14},
location = {Providence, Rhode Island, USA},
series = {STOC '85}
}

@article{apologies,
author = {Glikson, Ella and Asscher, Omri},
year = {2022},
month = {11},
pages = {107592},
title = {AI-mediated apology in a multilingual work context: Implications for perceived authenticity and willingness to forgive},
volume = {140},
journal = {Computers in Human Behavior},
doi = {10.1016/j.chb.2022.107592}
}

@inproceedings{10.5555/2343576.2343643,
author = {Kamar, Ece and Hacker, Severin and Horvitz, Eric},
title = {Combining human and machine intelligence in large-scale crowdsourcing},
year = {2012},
isbn = {0981738117},
publisher = {International Foundation for Autonomous Agents and Multiagent Systems},
address = {Richland, SC},
booktitle = {Proceedings of the 11th International Conference on Autonomous Agents and Multiagent Systems - Volume 1},
pages = {467–474},
numpages = {8},
location = {Valencia, Spain},
series = {AAMAS '12}
}

@inproceedings{10.5555/3061053.3061219,
author = {Kamar, Ece},
title = {Directions in hybrid intelligence: complementing AI systems with human intelligence},
year = {2016},
isbn = {9781577357704},
publisher = {AAAI Press},
booktitle = {Proceedings of the Twenty-Fifth International Joint Conference on Artificial Intelligence},
pages = {4070–4073},
numpages = {4},
location = {New York, New York, USA},
series = {IJCAI'16}
}

@inproceedings{Zhang_2020, series={FAT* ’20},
   title={Effect of confidence and explanation on accuracy and trust calibration in AI-assisted decision making},
   url={http://dx.doi.org/10.1145/3351095.3372852},
   DOI={10.1145/3351095.3372852},
   booktitle={Proceedings of the 2020 Conference on Fairness, Accountability, and Transparency},
   publisher={ACM},
   author={Zhang, Yunfeng and Liao, Q. Vera and Bellamy, Rachel K. E.},
   year={2020},
   month=jan, pages={295–305},
   collection={FAT* ’20} }

@article{Hui07022025,
author = {Emily Sein Yue Elim Hui},
title = {Incorporating Bloom’s taxonomy into promoting cognitive thinking mechanism in artificial intelligence-supported learning environments},
journal = {Interactive Learning Environments},
volume = {33},
number = {2},
pages = {1087--1100},
year = {2025},
publisher = {Routledge},
doi = {10.1080/10494820.2024.2364237},


URL = { 
    
        https://doi.org/10.1080/10494820.2024.2364237
    
    

},
eprint = { 
    
        https://doi.org/10.1080/10494820.2024.2364237
    
    

}

}

@article{driverassistance,
author = {Murtaza, Mohsin and Cheng, Chi-Tsun and Albahlal, Bader and Muslam, Muhana and Raza, Mansoor},
year = {2025},
month = {03},
pages = {},
title = {The impact of LLM chatbots on learning outcomes in advanced driver assistance systems education},
volume = {15},
journal = {Scientific Reports},
doi = {10.1038/s41598-025-91330-3}
}

@article{Tang2023,
author = "Ningzhi Tang and Meng Chen and Zheng Ning and Aakash Bansal and Yu Huang and Collin McMillan and Toby Jia-Jun Li",
title = "{An Empirical Study of Developer Behaviors for Validating and Repairing AI-Generated Code}",
year = "2023",
month = "3",
url = "https://kilthub.cmu.edu/articles/conference_contribution/An_Empirical_Study_of_Developer_Behaviors_for_Validating_and_Repairing_AI-Generated_Code/22223533",
doi = "10.1184/R1/22223533.v1"
}

@inproceedings{danry2024deceptiveaisystemsexplanations,
author = {Danry, Valdemar and Pataranutaporn, Pat and Groh, Matthew and Epstein, Ziv},
title = {Deceptive Explanations by Large Language Models Lead People to Change their Beliefs About Misinformation More Often than Honest Explanations},
year = {2025},
isbn = {9798400713941},
publisher = {Association for Computing Machinery},
address = {New York, NY, USA},
url = {https://doi.org/10.1145/3706598.3713408},
doi = {10.1145/3706598.3713408},
booktitle = {Proceedings of the 2025 CHI Conference on Human Factors in Computing Systems},
articleno = {933},
numpages = {31},
location = {
},
series = {CHI '25}
}

@misc{singh2025protectinghumancognitionage,
      title={Protecting Human Cognition in the Age of AI}, 
      author={Anjali Singh and Karan Taneja and Zhitong Guan and Avijit Ghosh},
      year={2025},
      eprint={2502.12447},
      archivePrefix={arXiv},
      primaryClass={cs.CY},
      url={https://arxiv.org/abs/2502.12447}, 
}

@book{kahneman2011thinking,
  address = {New York},
  author = {Kahneman, Daniel},
  description = {Thinking, Fast and Slow: Amazon.de: Daniel Kahneman: Englische Bücher},
  isbn = {9780374275631 0374275637},
  publisher = {Farrar, Straus and Giroux},
  refid = {706020998},
  title = {Thinking, fast and slow},
  url = {https://www.amazon.de/Thinking-Fast-Slow-Daniel-Kahneman/dp/0374275637/ref=wl_it_dp_o_pdT1_nS_nC?ie=UTF8&colid=151193SNGKJT9&coliid=I3OCESLZCVDFL7},
  year = 2011
}

@article{navon1979economy,
  title={On the economy of the human-processing system.},
  author={Navon, David and Gopher, Daniel},
  journal={Psychological review},
  volume={86},
  number={3},
  pages={214},
  year={1979},
  publisher={American Psychological Association}
}

@article{hallucinations,
author = {Alkaissi, Hussam and Mcfarlane, Samy},
year = {2023},
month = {02},
pages = {},
title = {Artificial Hallucinations in ChatGPT: Implications in Scientific Writing},
volume = {15},
journal = {Cureus},
doi = {10.7759/cureus.35179}
}

@inproceedings{ehsan2020humancenteredexplainableaireflective,
  title={Human-centered explainable ai: Towards a reflective sociotechnical approach},
  author={Ehsan, Upol and Riedl, Mark O},
  booktitle={International conference on human-computer interaction},
  pages={449--466},
  year={2020},
  organization={Springer}
}

@incollection{gergle2014experimental,
  title={Experimental research in HCI},
  author={Gergle, Darren and Tan, Desney S},
  booktitle={Ways of Knowing in HCI},
  pages={191--227},
  year={2014},
  publisher={Springer}
}

@inproceedings{allen2023hurry,
  title={In a Hurry: How Time Constraints and the Presentation of Web Search Results Affect User Behaviour and Experience},
  author={Allen, Garrett and Beijen, Mike and Maxwell, David and Gadiraju, Ujwal},
  booktitle={International Conference on Web Engineering},
  pages={221--235},
  year={2023},
  organization={Springer}
}

@inproceedings{qurashi2020document,
  title={Document processing: Methods for semantic text similarity analysis},
  author={Qurashi, Abdul Wahab and Holmes, Violeta and Johnson, Anju P},
  booktitle={2020 international conference on INnovations in Intelligent SysTems and Applications (INISTA)},
  pages={1--6},
  year={2020},
  organization={IEEE}
}

@inproceedings{vani2015investigating,
  title={Investigating the impact of combined similarity metrics and POS tagging in extrinsic text plagiarism detection system},
  author={Vani, K and Gupta, Deepa},
  booktitle={2015 international conference on advances in computing, communications and informatics (ICACCI)},
  pages={1578--1584},
  year={2015},
  organization={IEEE}
}

@article{chen_more_2025,
	title = {More {AI} {Assistance} {Reduces} {Cognitive} {Engagement}: {Examining} the {AI} {Assistance} {Dilemma} in {AI}-{Supported} {Note}-{Taking}},
	volume = {9},
	issn = {2573-0142},
	shorttitle = {More {AI} {Assistance} {Reduces} {Cognitive} {Engagement}},
	url = {https://dl.acm.org/doi/10.1145/3757632},
	doi = {10.1145/3757632},
	language = {en},
	number = {7},
	urldate = {2025-11-05},
	journal = {Proceedings of the ACM on Human-Computer Interaction},
	author = {Chen, Xinyue and Ruan, Kunlin and Ju, Kexin Phyllis and Yap, Nathan and Wang, Xu},
	month = oct,
	year = {2025},
	pages = {1--29},
}

@inproceedings{reimers2019sentencebertsentenceembeddingsusing,
  title={Sentence-bert: Sentence embeddings using siamese bert-networks},
  author={Reimers, Nils and Gurevych, Iryna},
  booktitle={Proceedings of the 2019 conference on empirical methods in natural language processing and the 9th international joint conference on natural language processing (EMNLP-IJCNLP)},
  pages={3982--3992},
  year={2019}
}

@misc{loria2026textblob,
  author       = {Steven Loria and contributors},
  title        = {TextBlob Documentation (Release 0.19.0)},
  year         = {2026},
  howpublished = {Read the Docs},
  note         = {Accessed 2026-01-06}
}

@article{jaccard1901,
  author  = {Jaccard, Paul},
  title   = {Etude comparative de la distribution florale dans une portion des Alpes et des Jura},
  journal = {Bulletin de la Societe Vaudoise des Sciences Naturelles},
  volume  = {37},
  pages   = {547--579},
  year    = {1901}
}

@article{zhao_als-mrs_2022,
	title = {{ALS}-{MRS}: {Incorporating} aspect-level sentiment for abstractive multi-review summarization},
	volume = {258},
	issn = {0950-7051},
	url = {https://www.sciencedirect.com/science/article/pii/S0950705122010358},
	doi = {https://doi.org/10.1016/j.knosys.2022.109942},
	journal = {Knowledge-Based Systems},
	author = {Zhao, Qingjuan and Niu, Jianwei and Liu, Xuefeng},
	year = {2022},
	pages = {109942},
}

@article{10.1145/3710946,
author = {de Jong, Sander and Paananen, Ville and Tag, Benjamin and van Berkel, Niels},
title = {Cognitive Forcing for Better Decision-Making: Reducing Overreliance on AI Systems Through Partial Explanations},
year = {2025},
issue_date = {May 2025},
publisher = {Association for Computing Machinery},
address = {New York, NY, USA},
volume = {9},
number = {2},
url = {https://doi.org/10.1145/3710946},
doi = {10.1145/3710946},
journal = {Proc. ACM Hum.-Comput. Interact.},
month = may,
articleno = {CSCW048},
numpages = {30}
}

@inproceedings{Lee_2022, series={CHI ’22},
   title={CoAuthor: Designing a Human-AI Collaborative Writing Dataset for Exploring Language Model Capabilities},
   url={http://dx.doi.org/10.1145/3491102.3502030},
   DOI={10.1145/3491102.3502030},
   booktitle={CHI Conference on Human Factors in Computing Systems},
   publisher={ACM},
   author={Lee, Mina and Liang, Percy and Yang, Qian},
   year={2022},
   month=apr, pages={1–19},
   collection={CHI ’22} }

@inproceedings{coenen2021wordcrafthumanaicollaborativeeditor,
author = {Yuan, Ann and Coenen, Andy and Reif, Emily and Ippolito, Daphne},
title = {Wordcraft: Story Writing With Large Language Models},
year = {2022},
isbn = {9781450391443},
publisher = {Association for Computing Machinery},
address = {New York, NY, USA},
url = {https://doi.org/10.1145/3490099.3511105},
doi = {10.1145/3490099.3511105},
booktitle = {Proceedings of the 27th International Conference on Intelligent User Interfaces},
pages = {841–852},
numpages = {12},
location = {Helsinki, Finland},
series = {IUI '22}
}

@article{10.1145/3449091,
author = {Mieczkowski, Hannah and Hancock, Jeffrey T. and Naaman, Mor and Jung, Malte and Hohenstein, Jess},
title = {AI-Mediated Communication: Language Use and Interpersonal Effects in a Referential Communication Task},
year = {2021},
issue_date = {April 2021},
publisher = {Association for Computing Machinery},
address = {New York, NY, USA},
volume = {5},
number = {CSCW1},
url = {https://doi.org/10.1145/3449091},
doi = {10.1145/3449091},
journal = {Proc. ACM Hum.-Comput. Interact.},
month = apr,
articleno = {17},
numpages = {14}
}

@inproceedings{10.1145/3544548.3581351,
author = {Fu, Liye and Newman, Benjamin and Jakesch, Maurice and Kreps, Sarah},
title = {Comparing Sentence-Level Suggestions to Message-Level Suggestions in AI-Mediated Communication},
year = {2023},
isbn = {9781450394215},
publisher = {Association for Computing Machinery},
address = {New York, NY, USA},
url = {https://doi.org/10.1145/3544548.3581351},
doi = {10.1145/3544548.3581351},
booktitle = {Proceedings of the 2023 CHI Conference on Human Factors in Computing Systems},
articleno = {103},
numpages = {13},
location = {Hamburg, Germany},
series = {CHI '23}
}

@inproceedings{10.1145/3701716.3717543,
author = {Sussman, Kristen and Carter, Daniel},
title = {Detecting Effects of AI-Mediated Communication on Language Complexity and Sentiment},
year = {2025},
isbn = {9798400713316},
publisher = {Association for Computing Machinery},
address = {New York, NY, USA},
url = {https://doi.org/10.1145/3701716.3717543},
doi = {10.1145/3701716.3717543},
booktitle = {Companion Proceedings of the ACM on Web Conference 2025},
pages = {2689–2693},
numpages = {5},
location = {Sydney NSW, Australia},
series = {WWW '25}
}

@inproceedings{10.1145/3630106.3658941,
author = {Kim, Sunnie S. Y. and Liao, Q. Vera and Vorvoreanu, Mihaela and Ballard, Stephanie and Vaughan, Jennifer Wortman},
title = {"I'm Not Sure, But...": Examining the Impact of Large Language Models' Uncertainty Expression on User Reliance and Trust},
year = {2024},
isbn = {9798400704505},
publisher = {Association for Computing Machinery},
address = {New York, NY, USA},
url = {https://doi.org/10.1145/3630106.3658941},
doi = {10.1145/3630106.3658941},
booktitle = {Proceedings of the 2024 ACM Conference on Fairness, Accountability, and Transparency},
pages = {822–835},
numpages = {14},
location = {Rio de Janeiro, Brazil},
series = {FAccT '24}
}

@inproceedings{10.1145/3630106.3658901,
author = {Guo, Ziyang and Wu, Yifan and Hartline, Jason D. and Hullman, Jessica},
title = {A Decision Theoretic Framework for Measuring AI Reliance},
year = {2024},
isbn = {9798400704505},
publisher = {Association for Computing Machinery},
address = {New York, NY, USA},
url = {https://doi.org/10.1145/3630106.3658901},
doi = {10.1145/3630106.3658901},
booktitle = {Proceedings of the 2024 ACM Conference on Fairness, Accountability, and Transparency},
pages = {221–236},
numpages = {16},
location = {Rio de Janeiro, Brazil},
series = {FAccT '24}
}

\appendix

\section{Task Prompts and AI Suggestions}
\label{appendix:tasks}

\subsection*{Task A - Analysis}

\paragraph{Prompt.}
Evaluate Bob's decision to wait at the old restaurant site for twenty years. Judge whether his choice was wise or misguided, using two story details. Finally, conclude what this reveals about the theme of friendship versus duty.

\paragraph{AI Suggestion.}
Bob’s decision to wait at the old restaurant site for twenty years reflects a mix of admirable loyalty and ultimately misguided hope. His declaration that he “came a thousand miles to stand here tonight” highlights his deep commitment to the promise he made, while his certainty that Jimmy “will never forget” shows his unwavering faith in his friend’s character. Yet the ironic moment when Bob realizes, “Twenty years is a long time, but not long enough to change the shape of a man’s nose,” reveals how his trust has blinded him to the possibility that their lives—and moral paths—have diverged. Together, these details show that Bob’s long wait, though rooted in sincere friendship, becomes a tragic misjudgment. Ultimately, the story suggests that the tension between friendship and duty can force painful choices, and Jimmy’s decision to prioritize justice over personal loyalty underscores that duty may demand sacrifices that even lifelong bonds cannot prevent.

\subsection*{Task B - Creative}

\paragraph{Prompt.}
Rewrite the final scene from Jimmy Wells's point of view (begin when he arrives on the street). Show how Jimmy processes the twist when he realizes Bob is a wanted man.

\paragraph{AI Suggestion.}
Suggested Structure:
\begin{enumerate}
  \item Introduction --- Jimmy's emotions as he arrives.
  \item Recognition --- His reaction upon seeing Bob.
  \item Conflict --- Realizing Bob is a wanted man.
  \item Resolution --- Choosing duty with compassion.
  \item Reflection --- Jimmy's lasting feelings.
\end{enumerate}

Suggested Focus:
Explore Jimmy's emotional conflict — his loyalty to friendship versus his duty as a policeman — and how he chooses a compassionate yet responsible path.

\end{document}